\documentclass[preprint,12pt]{elsarticle}

\biboptions{sort&compress}
\usepackage{amsmath}
\usepackage{amsfonts,amssymb}
\usepackage{graphicx,subfig}
\usepackage{color}
\usepackage{psfrag} %FOR DRAWING PS
\usepackage{epstopdf}	
\usepackage{mathrsfs,bbm}
\usepackage{lineno,hyperref}
\usepackage[section]{placeins}
\usepackage{anysize}
\journal{Physica A}

\begin{document}
\begin{frontmatter}

\title{Corrupted bifractal features in finite uncorrelated power-law distributed data}

\author[IFISC]{Felipe Olivares\corref{mycorrespondingauthor}}
\cortext[mycorrespondingauthor]{Corresponding author}
\ead{olivaresfe@gmail.com }

\author[IFISC]{Massimiliano Zanin}

\address[IFISC]{Instituto de F\'isica Interdisciplinar y Sistemas Complejos CSIC-UIB, Campus Universitat de les Illes Balears, E-07122 Palma de Mallorca, Spain.}

\begin{abstract}

Multifractal Detrended Fluctuation Analysis stands out as one of the most reliable methods for unveiling multifractal properties, specially when real-world time series are under analysis.
However, little is known about how several aspects, like artefacts during the data acquisition process, affect its results. In this work we have numerically investigated the performance of Multifractal Detrended Fluctuation Analysis applied to synthetic finite uncorrelated data following a power-law distribution in the presence of additive noise, and periodic and randomly-placed outliers. We have found that, on one hand, spurious multifractality is observed as a result of data finiteness, while additive noise leads to an underestimation of the exponents $h_q$ for $q<0$ even for low noise levels. On the other hand, additive periodic and randomly-located outliers result in a corrupted inverse multifractality around $q=0$. Moreover, the presence of randomly-placed outliers corrupts the entire multifractal spectrum, in a way proportional to their density. As an application, the multifractal properties of the time intervals between successive aircraft landings at three major European airports are investigated.
\end{abstract}


\end{frontmatter}

\section{Introduction}

Originated from the concept of multifractal sets \cite{benzi1984multifractal, chhabra1989direct,meneveau1989measurement}, multifractality, i.e. when the scaling becomes a local property rather than a global one, has been used for characterising the local scaling behaviour in empirical signals coming from numerous areas of science, as for instance air traffic flow~\cite{zhang2019multifractal}, financial data~\cite{cajueiro2007long,jin2006origins,zunino2008multifractal,zhang2019analysis,rak2018quantitative}, brain electrical activity~\cite{maity2015multifractal}, human heartbeat dynamics~\cite{ivanov1999multifractality}, extreme events in atmospheric turbulence~\cite{olivares2021high}, river discharge and precipitation~\cite{kantelhardt2003multifractality}, seismic records~\cite{telesca2005multifractal,telesca2006measuring} wind speed~\cite{laib2018multifractal}, musical signals~\cite{oswikecimka2011computational}, narrative texts~\cite{ausloos2012generalized,drozdz2016quantifying} and avian influenza~\cite{leung2011temporal}. As evidenced, many experimental measurements require multiple scaling exponents for a proper characterisation of the underlying complex dynamics. The multifractal nature originates from the presence of nonlinear correlations for small and large fluctuations~\cite{kantelhardt2002multifractal}. Nevertheless, a multifractal behaviour can also be observed due to broad probability distributions of the data~\cite{kantelhardt2002multifractal,rak2018quantitative} or to finite-size linearly long-term correlated time series~\cite{grech2013multifractal}. When analysing real-world data, it is common to face an interplay between these three sources of multifractality. Nevertheless, multifractality originated by nonlinear correlations is normally considered the only real one, placing linear correlations in short sequences and the heavy tailed distributed fluctuations as ingredients yielding ``spurious" multifractality~\cite{rak2018quantitative}. Particularly, power-law distributed data or distributions with tails obeying a power-law give rise to the most simple case of multifractality---bi-fractality~\cite{kantelhardt2002multifractal,nakao2000multi}---that is usually observed in systems exhibiting phase transitions~\citep{oswiecimka2006wavelet} and Levy processes~\citep{nakao2000multi}, as for instance financial markets~\citep{rak2018quantitative}. 

Among all multifractal approaches available in the literature~\cite{barabasi1991multifractal,muzy1994multifractal,kantelhardt2002multifractal,zhou2008multifractal,serrano2009wavelet,xiong2017weighted,jiang2019multifractal}, the most frequently used techniques to quantify the multi-scaling nature in synthetic and empirical data sequences are the Wavelet Transform Modulus Maxima method (WTMM)~\cite{muzy1994multifractal} and the Multifractal Detrended Fluctuation Analysis (MF-DFA)~\cite{kantelhardt2002multifractal}, the latter one being a generalisation of the classical DFA method~\cite{kantelhardt2001detecting}. Despite the fact that both methodologies can remove unwanted polynomial trends in the data, MF-DFA one stands out over the WTMM approach for being more accurate with shorter signals. Furthermore, it is more reliable in properly detecting monofractal and bifractal nature~\cite{oswiecimka2006wavelet}, and easier to implement~\cite{thompson2016multifractal,ihlen2012introduction}. In general, MF-DFA is recommended when the multifractal properties of the signal under analysis are unknown \emph{a priori}~\cite{oswiecimka2006wavelet}. Specifically, MF-DFA generalises the classical Hurst exponent  to a local exponent $h$ to unravel the heterogeneous scaling in the data. In such a manner,  it identifies and quantifies local scaling behaviours---multiple scaling exponents.

The identification and quantification of multifractal features in empirical data is a challenging task due to the unavoidable contamination with artefacts inherent to the data acquisition process~\cite{drozdz2010quantitative,turiel2006numerical,ludescher2011spurious,gulich2012effects,oswiecimka2020wavelet}. It is worth mentioning here that Ludescher \emph{et al.}~\cite{ludescher2011spurious} have studied the performance of the MF-DFA method for characterising mono and multifractal synthetic sequences under the influence of additive noise, short-term memory and periodicities. Particularly, they found that multifractal analysis can easily be  biased by those artefacts; for instance, multifractality is underestimated for multifractal signals with additive periodic trends. Later, following the same idea, Gulich \emph{et al.}~\cite{gulich2012effects} have studied the spurious multifractality generated by the influence of additive coloured noises in multifractal time series. Still further, it has been shown that a spurious broad multifractal spectrum is obtained when monofractal records having local isolated singularities are considered, such as records with outliers due to measurement errors~\cite{oswiecimka2020wavelet}. 

The aim of this work is to address how bifractal properties manifest in finite uncorrelated data obeying a power-law distribution, in the presence of some artefacts usually related to empirical recordings, by implementing MF-DFA methodology. More specifically, we focus on (i) finite size effects, since experimental data sets are constrained by data availability and stationarity, (ii) the influence of additive noise that is inherent to any measurement, and (iii) the presence of periodic and random outliers, both intrinsic to the underlying dynamics or due to errors in the data acquisition process.  Additionally, our findings are applied to the study of real-world time series describing the time between successive landings at three major European airports. 

\section{Multifractal Detrended Fluctuation Analysis (MF-DFA)}

The MF-DFA algorithm has been described in~\citep{kantelhardt2002multifractal}, and a detailed implementation can be found in~\cite{thompson2016multifractal,ihlen2012introduction}. Yet, for the sake of completeness, its main elements are described below. 

Given a time series $\mathcal{X}_{t} = \lbrace x_t, t = 1, ..., M \rbrace$, with $M$ being the number of observations, the cumulated time series $Y(i)=\sum_{t=1}^{i}(x_t - \langle x\rangle )$ is considered, where $\langle x\rangle$ stands for mean value. This profile is divided into $\lfloor N/s \rfloor$\footnote{$\lfloor c \rfloor$ denotes the largest integer less than or equal to $c$} non-overlapping windows of equal length $s$. A local polynomial fit $y_{\nu,m}(i)$ of degree $m$ is fitted to the profile of each window $\nu = 1, ..., \lfloor N/s \rfloor$. The degree of the polynomial can be varied to eliminate constant ($m=0$), linear ($m=1$), quadratic ($m=2$), or higher order trends of the profile.  The variance of the detrended time series is aftward evaluated by averaging over all data points $i$ in each segment $\nu$,
\begin{equation}
	F_{m}^{2}(s) =  \frac{1}{s} \sum_{i=1}^{s} \left\lbrace Y[(\nu -1)s + i] - y_{\nu,m}(i) \right\rbrace ^{2},
\end{equation}
for $\nu = 1, ..., \lfloor N/s \rfloor$. Aiming at analysing the influence of fluctuations of different magnitudes and at different time scales, the generalised $q$th order fluctuation function is defined by
\begin{equation}
	F_{q}(s) = \left\lbrace  \frac{1}{\lfloor N/s \rfloor} \sum_{i=1}^{\lfloor N/s \rfloor} [F_{m}^{2}(\nu,s)]^{q/2} \right\rbrace ^{1/q}.
\end{equation}
When $q=0$, a logarithmic averaging procedure has to be employed because of the divergent exponent 
\begin{equation}
	F_{0}(s) = \text{exp} \left\lbrace  \frac{1}{4\lfloor N/s \rfloor} \sum_{\nu=1}^{2\lfloor N/s \rfloor} \ln [F_{m}^{2}(\nu,s)]\right\rbrace.
\end{equation}
For $q=2$ the classical fractal DFA algorithm is retrieved~\cite{kantelhardt2001detecting}. Generally, for long-term power-law correlated data, it holds that
\begin{equation}
	F_{q}(s) \sim s^{h_{q}},
\end{equation}
inside a certain range of $s$. The scaling exponents $h_{q}$ are usually known as generalised Hurst exponents, and allow accounting for heterogeneous scaling. Ideally, for a monofractal time series, $h_q$ is independent of $q$ and equal to the Hurst exponent, $H$. Its value quantifies the degree of correlation in the data: if $H=0.5$ the time series is uncorrelated, while $H>0.5$ indicates long-term correlations or persistence. On the other hand, antipersistent behaviour is characterised by $H<0.5$. 

A bi- or multifractal structure is observed when the scaling behaviours of small and large fluctuations are different. In this case $h_q$ decreases with $q$, and the main Hurst exponent can be estimated from the second moment ($h_2 = H$). More specifically, the generalised Hurst exponent with negative order $q$ describes the scalings of small fluctuations, because the segments $\nu$ with small variance dominate the average $F_{q}$. On the contrary, for positive orders $q$, the windows $\nu$ with large variance have a stronger influence, and thus $h_q$ focuses on large fluctuations. The strength of the multifractality present in the data is usually defined as the spread of the generalised Hurst exponent~\cite{grech2013multifractal}. As small fluctuations are characterised by larger scaling exponents than those associated with large fluctuations, the multifractality degree can be quantified by
\begin{equation}
	\Delta h_{q} \equiv h(q_{\text{min}}) - h(q_{\text{max}}),
	\label{Eq_MF_degree}
\end{equation}
where $q_{\text{min}}$ and $q_{\text{max}}$ are respectively the minimal and maximal value of the moment $q$ considered in the analysis. Normally $q_{\text{min}}=-q_{\text{max}}$ is used. Another way to quantify multifractal features is by measuring the width of the singularity spectrum~\citep{halsey1986fractal,kantelhardt2002multifractal}. For further details about this methodology and its implementation see~\cite{thompson2016multifractal}. Specifically for a MATLAB implementation we recommend  Ref.~\cite{ihlen2012introduction}.

\section{Numerical analysis}\label{sec:num}

\subsection{A model of bi-fractal spectra}

We consider finite uncorrelated time series with power-law distribution function:
\begin{equation} 
	P(x) = x_{\text{min}} \, x^{-(\alpha +1)} \,\,\,\, \text{for} \,\,\, x_{\text{min}} \leq x < \infty,
	\label{Eq_power_law_pdf}
\end{equation}
with $\alpha>0$, and $x_{\text{min}}$ corresponding to the smallest value of $x$ for which the power law holds~\cite{newman2005power}. In the present study we set $P(x)=0$ for $x<x_{\text{min}}$~\cite{kantelhardt2001detecting}. Yet, empirical data normally do not follow a power-law over their entire range; in fact, the distribution usually deviates from the power-law from below the minimum value $x_{\text{min}}$~\cite{newman2005power}. For $\alpha \leq 2$, time series distributed according to Eq.~\ref{Eq_power_law_pdf} exhibit  multifractal scaling behaviours on all scales~\citep{kantelhardt2002multifractal}. It was shown in~\cite{kantelhardt2001detecting} that the generalised Hurst exponent for data distributed according to Eq.~\ref{Eq_power_law_pdf} can be expressed as 
\begin{equation}
h_q = \left\{
        \begin{array}{ll}
            1/\alpha & \quad q \leq \alpha, \\
            1/q & \quad q > \alpha,
        \end{array}
    \right.
    \label{Eq_h_q}
\end{equation}
which describes a bi-fractal nature, \emph{i.e.} a monofractal behavior ($h_q=$ constant) for $q \leq \alpha$, while a multifractal nature for $q>\alpha$~\citep{kantelhardt2002multifractal}. 

\subsection{Synthetic data generation}

For generating numerical uncorrelated sequences distributed according to a power-law we consider a transformation procedure. Being $r_i$ random real numbers uniformly distributed in the interval $[0,1]$, these are transformed according to $r_i \rightarrow x_i=x_{\text{min}}\,r_i^{-1/\alpha}$, to obtain random power-law distributed real numbers $x_i$ in the range $[x_{\text{min}},\infty)$~\citep{kantelhardt2002multifractal}. We have generated a set of one hundred independent realisations of length $N= 10^{n}$ with $n={3,4,5,6}$, different values of $\alpha \in [0.5,2]$, and $x_{\text{min}}=1$. The MF-DFA method considers 30 time scales $s \in [10,N/10]$ equally distributed in a logarithmic scale. We set the moments $q \in [-10,10]$ with a step equal to 0.1. Results shown in the present work were obtained by using a detrending polynomial of second order ($m=2$). Quantities averaged over one hundred realisations are reported hereafter. 

\subsection{Finite-size effects}\label{sec:size}

\begin{figure}[!tb]
\centering\includegraphics[width=\linewidth]{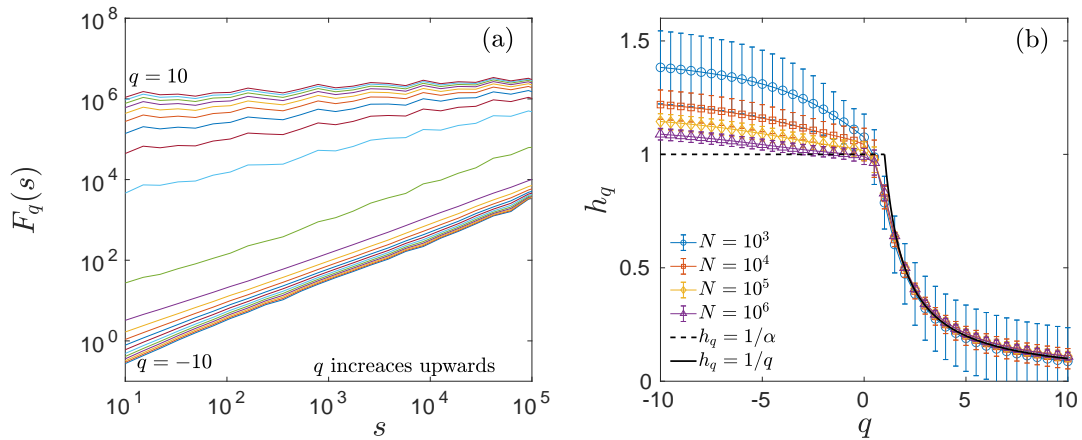}
\caption{(a) Generalised Fluctuation function as a function of the time scale $s$ with $q \in [-10,10]$ for $\alpha = 1$ and $N=10^{6}$ (only integer values of $q$ are depicted for sake of clarity). (b) Generalised Hurst exponent $h_q$ a function of $q$ for the same value of $\alpha$ and for different values of the time series length $N$. Dashed and solid lines describe the theoretical curve for power-law distributed data with $\alpha=1$ at long scales for $q \leq \alpha$ and $q>\alpha$ respectively.}
\label{fig:num1}
\end{figure}

\begin{figure}[!tb]
\centering\includegraphics[width=\linewidth]{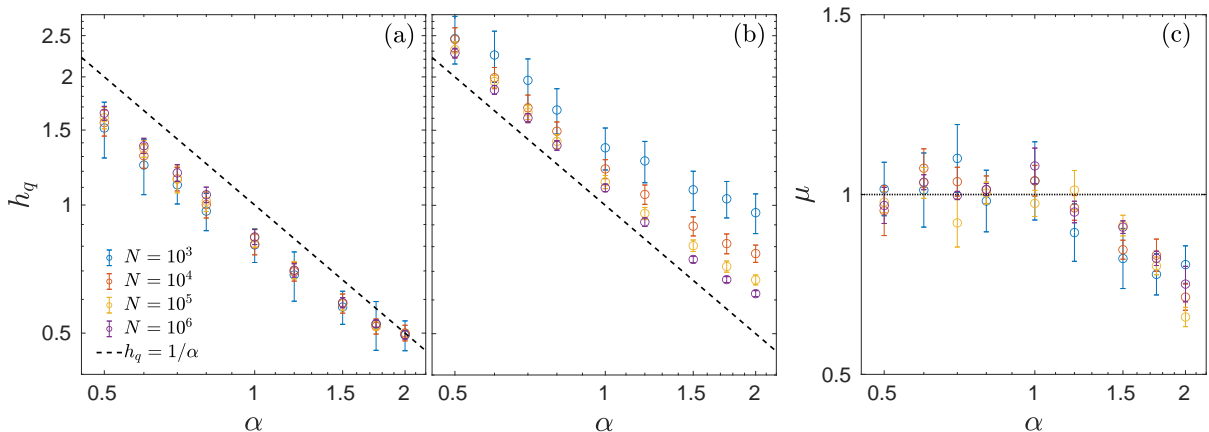}
\caption{(a) Generalised Hurst exponent for $q=\alpha$ as a function of $\alpha$ for different values of $N$. (b) same as (a) but with $h_q$ evaluated in $q=-10$. The dashed black lines in both previous panels represent the theoretical expected power law $h_q=1/\alpha$. (c) Exponent $\mu$ as a function of $\alpha$ for different values of $N$.}
\label{fig:num2}
\end{figure}

We firstly investigate the validity of the theoretical model for $h_q$, given by Eq.~\ref{Eq_h_q}, for short time series. Fig.~\ref{fig:num1}(a) shows the generalised fluctuation function for $\alpha = 1$ and $N=10^{6}$. Only integer values of the moment $q$ are depicted. A good scaling behaviour is observed over the whole range of temporal scales considered and for all moments $q$. The slope of the $F_{q}(s)$ in the log-log scale decreases with $q$. These results qualitatively hold for all $\alpha \in [0.5,2]$. The slopes $h_q$ are depicted in Fig.~\ref{fig:num1}(b) as a function of $q$, as estimated by a linear least-squares fit and for different time series length. On one hand, we found a good agreement with the numerical simulation for $q>\alpha=1$, even for short time series. On the other hand, for $q<\alpha$ the theoretical monofractal behaviour is corrupted, mimicking a multifractal behaviour, i.e. $h_q$ decreasing with $q$. Fig.~\ref{fig:num2}(a)-(b) illustrate this spurious multifractality for different values of $\alpha$ and $q$. For $q=	\alpha$, the exponent $h_q$ is underestimated in comparison with the theoretical prediction, for all lengths $N$ considered in this analysis. Yet, for negative moments $q$, $h_q$ approaches from above as the time series length increases---see Fig.~\ref{fig:num2}(b). Note that the convergence to the scaling $h_q=1/\alpha$ with $N$ is slower as $\alpha$ grows.

Furthermore, with the purpose of characterising the multifractal behaviour for $q>\alpha$, we define
\begin{equation}
	h_q \sim q^{-\mu},
\end{equation}
where the exponent $\mu \equiv \mu(N,\alpha)$ accounts for deviations from the theoretical prediction ($\mu=1$). Fig.~\ref{fig:num2}(c) shows the estimated exponents $\mu$ as a function of $\alpha$. We found that, independently of the length $N$, the theoretical expected value is recovered up to $\alpha=1$. Yet, the exponent $\mu$ decreases for $\alpha \in [1,2]$.

%%%%%%%%%%%%%%%%%%%%%%%%%%%% FIG 1

%%%%%%%%%%%%%%%%%%%%%%%%%%%%%%%%%%%%%%%%%%%%%

%%%%%%%%%%%%%%%%%%%%%%%%%%%% FIG 2
%%%%%%%%%%%%%%%%%%%%%%%%%%%%%%%%%%%%%%%%%%%%%

\subsection{Influence of additive noise}\label{sec:noise}

We next study the influence of additive noise. For that, we define a noisy finite sequence as follows:
\begin{equation} 
	y_i = x_i + A\,\eta_i,
\end{equation}\label{Eq:noise}
where $x_i$ is a finite random time series distributed according to Eq.~\ref{Eq_power_law_pdf}, $\eta_i$ is a Gaussian white noise with zero mean and unit variance, and $A_n$ is the amplitud of the noise contamination defined as a multiple of the standard deviation of the original sequence $A = A_n \sigma_{x_i}$ ($A_n \in \mathbb{R}$). For generating the numerical time series we set the length $N=10^{5}$.

\begin{figure}[!tb]
\centering\includegraphics[width=\linewidth]{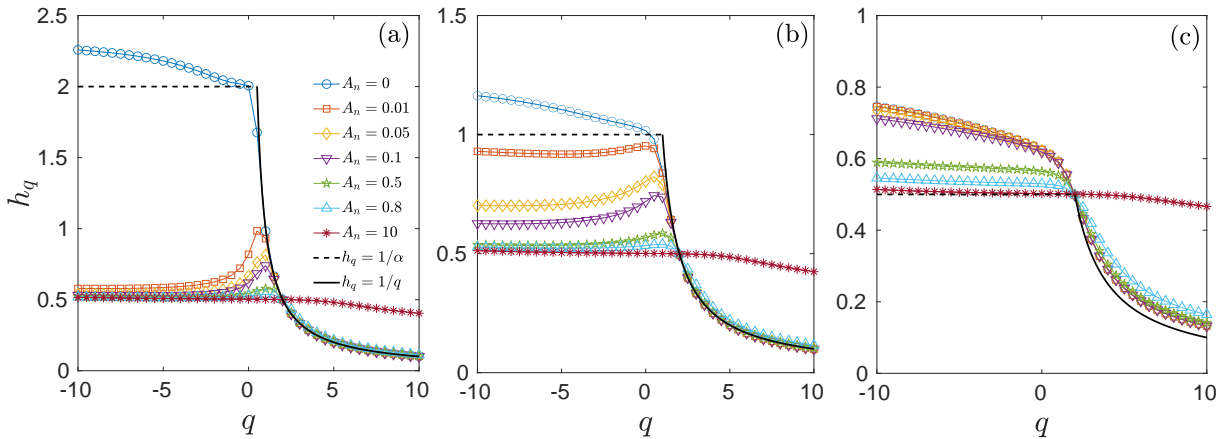}
\caption{Generalised Hurst exponent as a function of $q$ for different values of the noise intensity $A_n$ for (a) $\alpha=0.5$, (b) $\alpha=1$ and (c) $\alpha=2$. No error bars are depicted for the sake of clarity.}
\label{fig:num3}
\end{figure}

The generalised Hurst exponents are depicted in Fig.~\ref{fig:num3}, as a function of the moment $q$, for $\alpha=(0.5, 1, 2)$, and considering several intensities of additive noise. It stands out that, for $q<2$, the exponents $h_q$ are considerable underestimated, even for small amount of noise---negative values of the moment $q$ describes small fluctuations, which are closer to the noise level. Of course, for very large noise levels, as for instance for $A_n=10$, the exponent $h_q$ becomes constant and equal to 0.5, since the monofractal noise dominates the dynamics. This behaviour was previously observed in numerical simulations of noisy multifractal cascades~\cite{ludescher2011spurious,gulich2012effects}. Interestingly enough, additive noise offsets the spurious multifractal behaviour due to finite size effects as $\alpha \rightarrow 2$ (see Fig.~\ref{fig:num3}(c)), since $h_{q<\alpha} \rightarrow 0.5$. These two biases compete against each other and could eventually cancel out, e.g. when having short sequences in a noisy environment. Nevertheless, the multifractal degree is, without a doubt, underestimated for $\alpha<1$.

%%%%%%%%%%%%%%%%%%%%%%%%%%%% FIG 3

%%%%%%%%%%%%%%%%%%%%%%%%%%%%%%%%%%%%%%%%%%%%%

\subsection{Influence of additive periodic outliers}\label{sec:spikes}

We consider the influence of periodic outliers by defining:
\begin{equation} 
	y_i = x_i + I\,\delta_{\tau}(i), \,\,\,\,\,  \text{with} \,\,\,\delta_{\tau}(i) = \sum^{-\infty}_ {i = \infty} \delta(t-i\tau),
\end{equation}\label{Eq:noise}
where the original sequence $x_i$ is perturbed by a periodic spike characterised by the intensity $I$, defined in terms of the standard deviation of $x_i$,  $I = A_p \sigma_{x_i} +\eta_i$ ($A_p \in \mathbb{R}$ and $\eta_{i}$ is a Gaussian white noise with zero mean and unit variance), and with $\delta$ being the Dirac delta function. For generating the numerical time series we set the length $N=10^{5}$ and the period $\tau = 720$, which can represent, for instance, monthly spikes in hourly sampled data. 

\begin{figure}[!tb]
\centering\includegraphics[width=\linewidth]{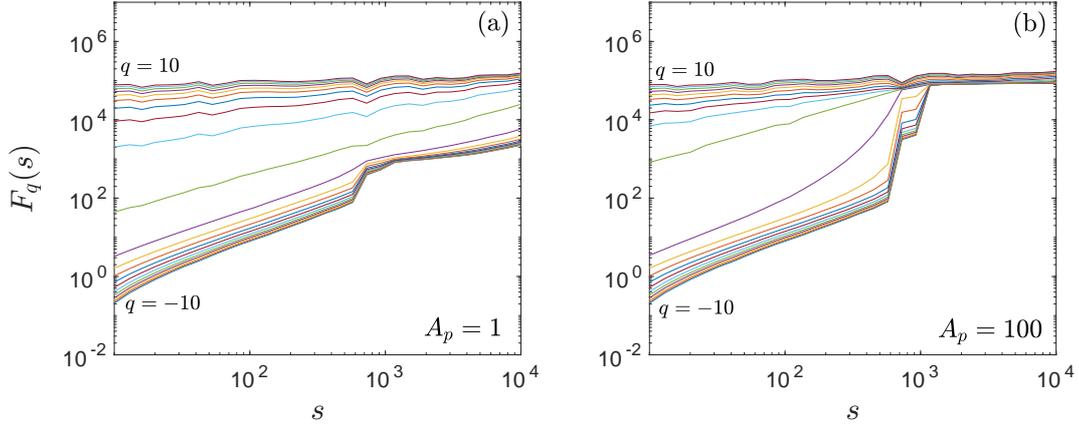}
\caption{Generalised fluctuation function as a function of the time scale $s$ with $q \in [-10,10]$ for $\alpha = 1$, $N=10^{5}$, for (a) $A_p = 1$ and (b) $A_p = 100$. Only integer values of $q$ are depicted for the sake of clarity.}
\label{fig:num4}
\end{figure}

\begin{figure}[!h]
\centering\includegraphics[width=\linewidth]{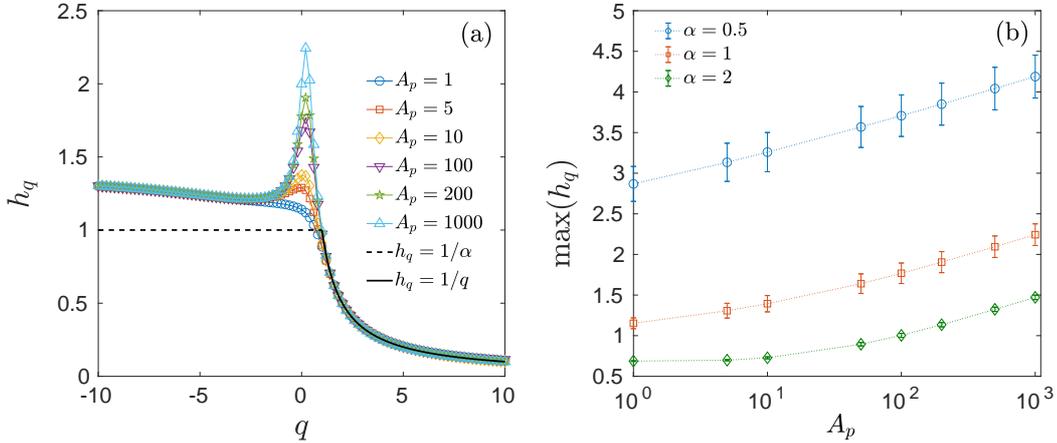}
\caption{(a) Generalised Hurst exponent as a function of $q$, estimated in the range $s \in [20,300]$, for $\alpha = 1$ $\tau=720$ and different values of $A_p$; and (b) maximum value of $h_q$ around $q = 0$ as a function of $A_p$. In panel (a) no error bars are depicted for the sake of clarity.}
\label{fig:num5}
\end{figure}

Figure~\ref{fig:num4} shows $F_q(s)$ as a function of the time scale $s$, for $\alpha = 1$ and for intensities of the periodic outliers (a) $A_p=1$ and (b) $A_p=100$. The larger the intensity of the outlier event, the larger the perturbation of the generalised fluctuation functions for $s>\tau$. However, we observe that it is still feasible to estimate of the slope $h_q$ if we restrict the fitting procedure to short time scales only, e.g. to $s<300$ data points. To illustrate, the estimated slopes $h_q$ are depicted in Fig.~\ref{fig:num5}(a). We found that the multifractal nature is corrupted around $q = 0$. The exponent $h_q$ is overestimated as $A_p$ increases, since large events interfere with the moments near the mean value. In synthesis, a spurious inverse multifractality is observed up to $q=0$. Yet, for $q>\alpha$, the decreasing behaviour of $h_q$ with $q$ remains the same. Consequently, the multifractal degree estimated by Eq.~\ref{Eq_MF_degree} is not dramatically affected. Same conclusions can be drawn for other values of $\alpha$. 

\begin{figure}[!h]
\centering\includegraphics[width=\linewidth]{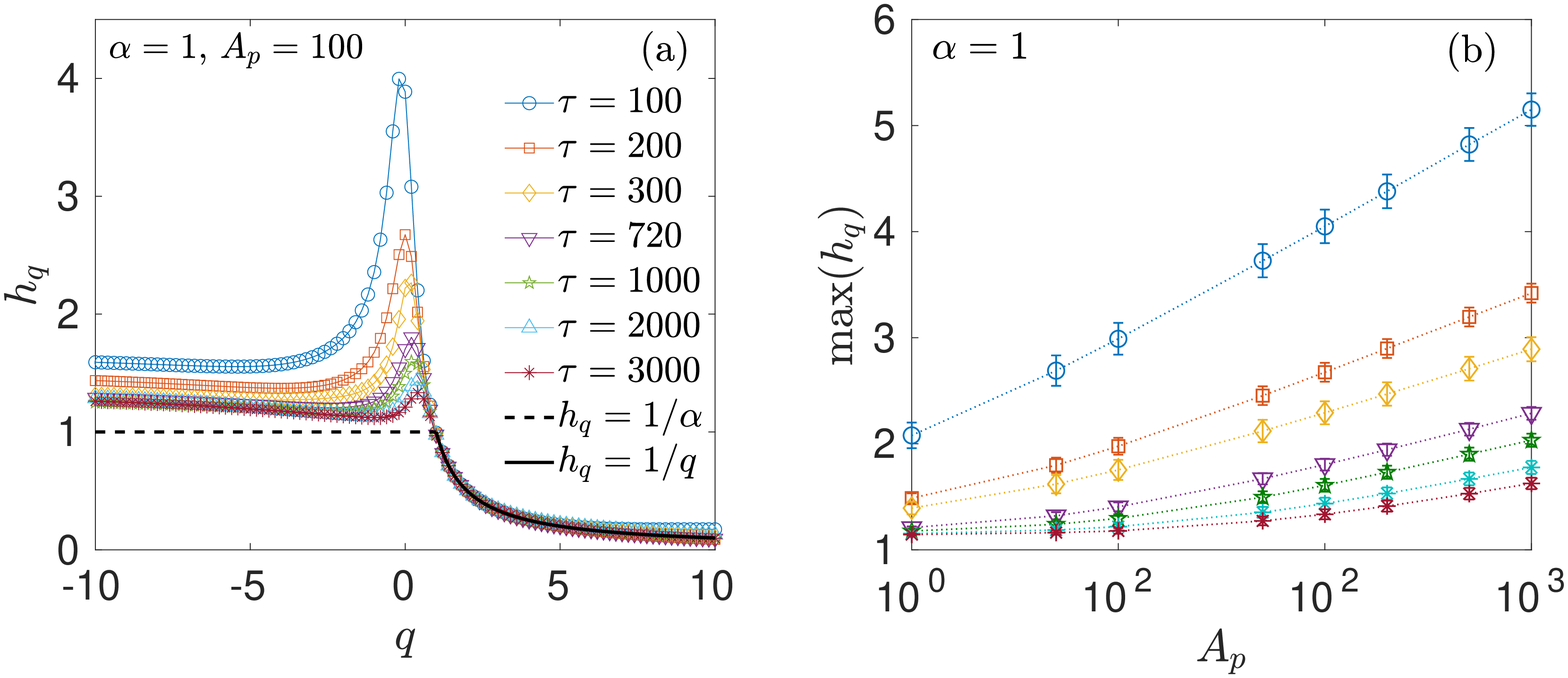}
\caption{(a) Generalised Hurst exponent as a function of $q$ for $\alpha = 1$, $A_p=100$ and different values of $\tau$ (no error bars are depicted for the sake of clarity); and (b) maximum value of $h_q$ around $q = 0$ as a function of $A_p$.}
\label{fig:num51}
\end{figure}

The maximum value reached by $h_q$ gives account, in some way, of the intensity of the spikes added to the sequence, as can be seen in Fig.~\ref{fig:num5}(b). As $\alpha$ increases, the overestimation of $h_q$ for low $A_p$ is compensated. Undeniably, when the intensity of the event is sufficiently high,  the estimated exponents $h_q$ are always overvalued, independently of $\alpha$. A similar behaviour is observed when different periodicities are considered. Fig.~\ref{fig:num51}(a) summarises the results for $\alpha=1$ and different values of $\tau$. The same range of temporal scales ($s \in [20,100]$) for all $\tau$ were considered to estimate $h_q$. On one hand, as the period increases, for a fixed intensity $A_p$, the inverse corrupted multifractality around $q=0$ is magnified. On the other hand, independently of the period, the maximum value of $h_q$ around $q=0$ linearly scales with the logarithm of $A_p$ (for large intensities), as can be seen in Fig.~\ref{fig:num51}(b).

%%%%%%%%%%%%%%%%%%%%%%%%%%%% FIG 4

%%%%%%%%%%%%%%%%%%%%%%%%%%%%%%%%%%%%%%%%%%%%%  

%%%%%%%%%%%%%%%%%%%%%%%%%%%% FIG 5

%%%%%%%%%%%%%%%%%%%%%%%%%%%%%%%%%%%%%%%%%%%%%

%%%%%%%%%%%%%%%%%%%%%%%%%%%% FIG 5.1

%%%%%%%%%%%%%%%%%%%%%%%%%%%%%%%%%%%%%%%%%%%%%

\subsection{Influence of randomly located outliers}\label{sec:random}

\begin{figure}[!h]
\centering\includegraphics[width=\linewidth]{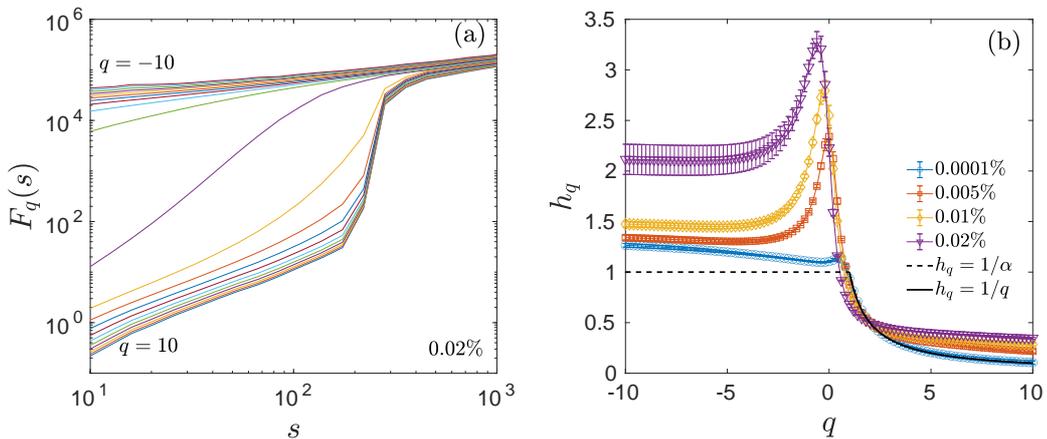}
\caption{(a) Generalised fluctuation function as a function of the time scale $s$ with $q \in [-10,10]$ for $\alpha = 1$, $N=10^{5}$, and 0.02$\%$ of the data being random outliers with an intensity $A_p=100$ (only integer values of $q$ are depicted for the sake of clarity). (b) Generalised Hurst exponent as a function of $q$ for $\alpha = 1$, $A_p=100$, and different percentages of randomly located events.}
\label{fig:num6}
\end{figure}

Lastly, we examine the effects when randomly located outliers perturb the original sequence $x_i$. Fig.~\ref{fig:num6} summarises the results for sequences of length $N=10^{5}$, characterised by $\alpha=1$ and perturbed with randomly located outliers with an intensity $A_p=100$. We found that the linear scaling of the function $F_q$ with $s$ is corrupted in a qualitatively similar way as though periodic spikes were perturbing the sequence, as can be seen in Fig.~\ref{fig:num6}(a). It is observed that even for a small amount of perturbing events, e.g. 0.02$\%$ of the data, the multifractal spectrum is dramatically squeezed upwards at large temporal scales.   

Figure~\ref{fig:num6}(b) contrasts the exponents $h_q$ estimated from sequences with different percentages of the data being randomly located outliers. For negative moments, there exists an overestimation of the exponents $h_q$, proportional to the amount of events in the data. On the other hand, for negative values of $q$, the theoretical decay $\sim q^{-1}$ is not longer valid. Moreover, not even the model $\sim q^{-\mu}$ can be fitted. Consequently, the entire multifractal spectrum is corrupted, and thus, the multifractal degree estimated by Eq.~\ref{Eq_MF_degree}. Not least, the peak of $h_q$ appreciated around $q \in 0$, accounts for the presence of outliers as aforementioned. 

%%%%%%%%%%%%%%%%%%%%%%%%%%%% FIG 6

%%%%%%%%%%%%%%%%%%%%%%%%%%%%%%%%%%%%%%%%%%%%%

\section{Application to landing return interval dynamics}

In order to show how these results can impact real-world analyses, we study time series describing aircraft landing dynamics. More specifically, we estimate the time between consecutive landings--- landing return interval---starting on May 1 of 2018 and ending on December 31 of 2019 (609 days) from Frankfurt, Heathrow and Tegel airports. Estimated landing times were extracted from ADS-B position reports, obtained from the OpenSky Network (https://opensky-network.org)~\cite{schafer2014bringing}. ADS-B (Automatic Dependent Surveillance - Broadcast) is a technology allowing aircraft to continuously send radio messages, stating their position and other information of relevance~\cite{williams2009gps,salcido2017analysis}; these messages are then received by ground stations, and integrated into coherent reports. For each airport, flights have been identified as performing a landing when the last known position was within a radius of 3 nautical miles from the center of the airport, and the last reported altitude below 500 meters. Note that there exists observational noise as a result of the uncertainty associated with the time of landing.

Tab.~\ref{Tab_1} lists, in the second column, the average number of flights per day. Additionally, being these airports close to populated areas, operations are not allowed at night (except for emergencies and other specified exceptions); the third column then reports the duration of this night inactivity period, as extracted from the corresponding Jeppesen's airport charts. Depending of the airport activity, different length $N$ of the landing return interval are obtained. The fourth column in Tab.~\ref{Tab_1} shows $N$ for each airport, being Tegel airport the one with the least number of landings per day and with the longest inactivity period.

\begin{table}[h!]
	\centering
	\caption{Main characteristics of the considered airports. First column: Airports names. Second column: night inactivity period in hours. Third column: average number of flights per day. Fourth column: time series lengths.}
    	\begin{tabular}{l| c c c }
     		Airports &  Night inactivity period (hrs.) & Averaged $\#$flights/day &$N$ \\ 
    		\hline \hline 
Frankfurt	& 6	&	740	&	451534	\\		 
Heathrow	& 6.5	 &	531	&	 323766	\\
Tegel		& 	7  &	79	&	 48112	\\
 			\hline     
       \end{tabular}
       \label{Tab_1}
\end{table}

\begin{figure*}[!th]
\centering\includegraphics[width=\linewidth]{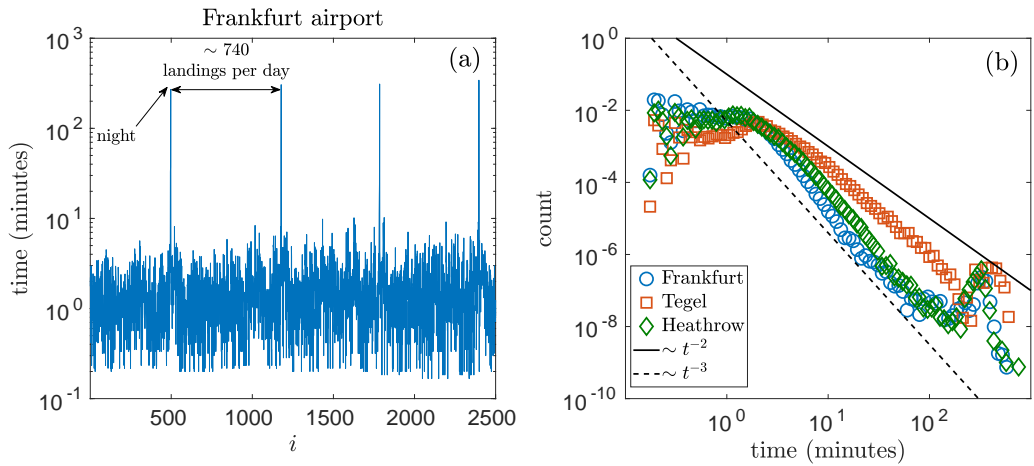}
\caption{(a) A portion of the landing return intervals from Frankfurt airport in semilog scale. (b) Histogram of the landing return interval for Frankfurt (solid blue circles), Tegel (solid red squares) and Heathrow (solid green diamonds) airports. Dotted and dashed black lines indicate power law $t^{-2}$ and $t^{-3}$ respectively.}
\label{fig:series_pdf_exp}
\end{figure*}

As a representative time series, in Fig.~\ref{fig:series_pdf_exp}(a) we illustrate a small portion of the landing return interval sequence from Frankfurt airport. Note that the vertical axis is in logarithmic scale for better visualisation of the outliers, corresponding to the inactivity time during the night. Figure~\ref{fig:series_pdf_exp}(b) shows the histograms, by using logarithmic binning~\citep{virkar2012power}, of the return interval sequences from the three airports. It is observed that a power-law scaling applies only for times larger than $t_{\text{min}} \sim 2$ and less than 100 minutes. We can therefore say that the distributions follow a power-law upon a certain range~\citep{newman2005power}. For comparison purposes, we have depicted a power-law scaling of $t^{-2}$ and $t^{-3}$---see dotted and dashed black lines in Fig.~\ref{fig:series_pdf_exp}(b), respectively. By a simple linear least-squares fit in the range $(2,100)$ minutes we have estimated the power-law parameter $\alpha$, reported in the second column in Tab.~\ref{Tab_2}. We found a correlation between the total number of flights per day and the distribution parameter $\alpha$. The more flights, the shorter the separation time between them, and consequently, the lower the probability of finding two flights widely separated in time. For times larger than approximately 100 minutes, a peak is observed, corresponding to the night inactivity period. Bearing in mind the noisy environment inherent to the data acquisition procedure, and furthermore, that inactivity times may be considered as outliers of the power-law distribution, these empirical sequences are ideal to study how these artefacts may corrupt their multiscaling properties.

%%%%%%%%%%%%%%%%%%%%%%%%%%%%% FIG 1 exp
%%%%%%%%%%%%%%%%%%%%%%%%%%%%%%%%%%%%%%%%%%%%%

\begin{table}[h!]
	\centering
	\caption{Multifractal properties of the considered airports. First column: Airports names. Second column: $\alpha$ value estimated by a linear fitting from the histogram in the range time $\in [2,100]$. Third column: Hurst exponents $h_2=H$. Fourth column: $\mu$ exponent estimated in the range $q \in [2,5]$. SD stands for the error from the fitting procedure.}
    	\begin{tabular}{c| c c c }

Airports & $\alpha \, \pm\, \text{SD} $ & $H \, \pm\, \text{SD} $ & $\mu \, \pm\, \text{SD} $  \\ 
    		\hline \hline 
Frankfurt& 2.3$\pm$0.1	&	0.53$\pm$0.01	&	0.90$\pm$0.005	\\	
Heathrow& 2$\pm$0.1	 &	0.58$\pm$0.02	&	 0.76$\pm$0.010\\
Tegel	& 	1.2$\pm$0.2  &	0.53$\pm$0.02	&	 0.56$\pm$0.017\\

 			\hline     
       \end{tabular}
       \label{Tab_2}
\end{table}

\begin{figure*}[!tb]
\centering\includegraphics[width=\linewidth]{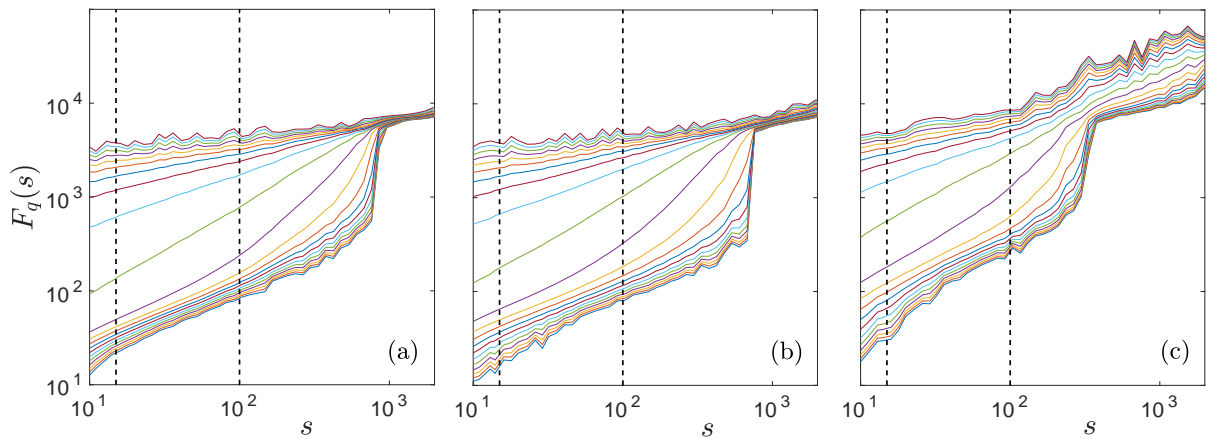}
\caption{$F_{q}$ as a function of $s$ for the landing return interval for (a) Frankfurt, (b) Heathrow, and (c) Tegel airports. Black dashed lines indicate the temporal range to estimate the exponents $h_q$. Only integer values of the moments $q$ used to estimate the generalised fluctuation function are depicted.}
\label{fig:Fq_exp}
\end{figure*}

We have analysed the multifractal properties of the landing return intervals with a second order polynomial. Sixty temporal scales $s \in [10,10^{4}]$ equally distributed in the logarithmic scale were considered and $q \in [-5,5]$ with a step equal to 0.1. Fig.~\ref{fig:Fq_exp} shows the $F_q$ as a function of $s$ for the landing return intervals for each airport. The time between the closing and opening hours squeezes upwards the scaling for a windows size $s>1000$ for Frankfurt and Heathrow airports. On the other hand, since Tegel has less flights per day, the scaling is corrupted at smaller time scales. These results show that the night inactivity period is having the role of a periodic outlier.

\begin{figure*}[!tb]
\centering\includegraphics[width=\linewidth]{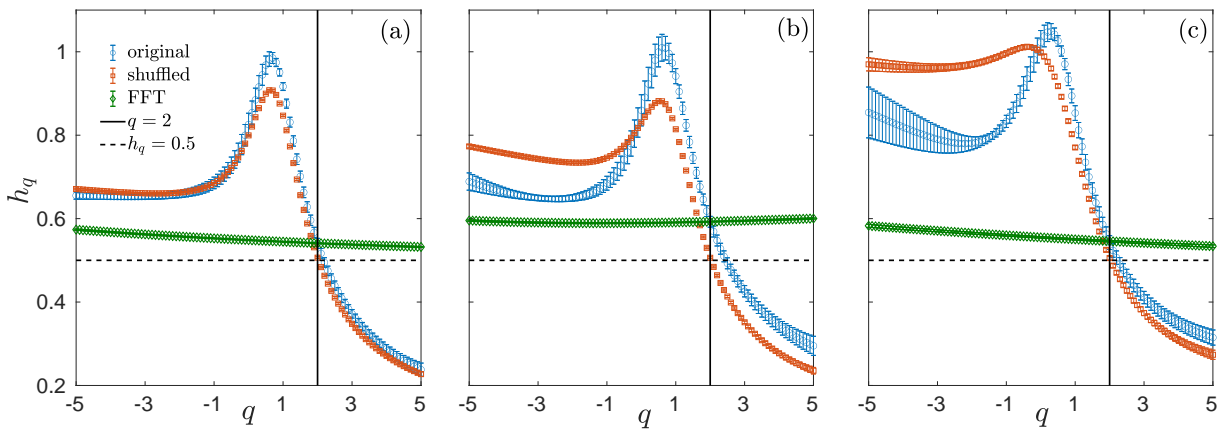}
\caption{Generalised Husrt exponents as a function of $q$ for (a) Frankfurt, (b) Heathrow and (c) Tegel airports. Error bars stand for the error from the fitting procedure for the original sequences. Solid black lines indicate the second moment $q=2$, and therefore the classical Hurst exponent $h_2=H$. Dashed black lines show $h=0.5$ (totally uncorrelated dynamics).}
\label{fig:hq_exp}
\end{figure*}

By setting a fitting range $s\in[15,100]$ for the three airports, we have estimated $h_q$ for all values of $q$. These results are depicted in Fig.~\ref{fig:hq_exp} (blue open circles). Slightly linear correlations are found, which are quantified by the classical Hurst exponent---see the intersection between solid vertical black line and $h_q$ in Fig.~\ref{fig:hq_exp}. The second column of Tab.~\ref{Tab_2} summarises the values of $H=h_2$ for each airport. The generalised exponents $h_q$ are overestimated in the range $q\in[0,1]$, most probably because of the long return times, i.e. of the nights. Out of this region, the evolution of $h_q$ behaves as one could expect from the probability distribution of the data, that is, bi-fractal. We have estimated the exponent $\mu$ in the range  $q \in [2,5]$---see last column in Tab.~\ref{Tab_2}. These results are in accordance with those obtained by using synthetic data (compare with Fig.~\ref{fig:num2}(c)).

In order to clarify the origin of the observed multifractality, we have contrasted the evolution of $h_q$ estimated from the original data, with what was obtained from surrogate data. By shuffling the data points, all temporal correlations are removed, and only the multifractal nature due to the distribution probability remains. Open red squares in Fig.~\ref{fig:hq_exp} show these results. One can observe that the multifractality is slightly affected for Frankfurt and Heathrow airports, indicating that the corrupted bi-fractality is mostly originated by the probability distribution. On the contrary, the result for Tegel airport shows some differences for negative moments. We hypothesise that this could be a product of the presence of nonlinearities in the dynamics. Certainly, a Hurst exponent equal to 0.5 is estimated for the shuffled versions of all the three sequences. Finally, by considering a Fourier Transform surrogate approach~\cite{rak2018quantitative} to retain all temporal correlations but changing the probability distribution to a Gaussian one by phase randomisation~\cite{theiler1992testing}, we found that the exponents $h_q$ are constant in the range $q \in [-5,5]$ and close to the Hurst exponent estimated from the original sequences---see open green diamonds in Fig.~\ref{fig:hq_exp}. These results confirm the origin of the bi-fractality. 

%%%%%%%%%%%%%%%%%%%%%%%%%%%%% FIG 2 exp
%%%%%%%%%%%%%%%%%%%%%%%%%%%%%%%%%%%%%%%%%%%%%

%%%%%%%%%%%%%%%%%%%%%%%%%%%%% FIG 3 exp
%%%%%%%%%%%%%%%%%%%%%%%%%%%%%%%%%%%%%%%%%%%%%

\begin{figure*}[!tb]
\centering\includegraphics[width=\linewidth]{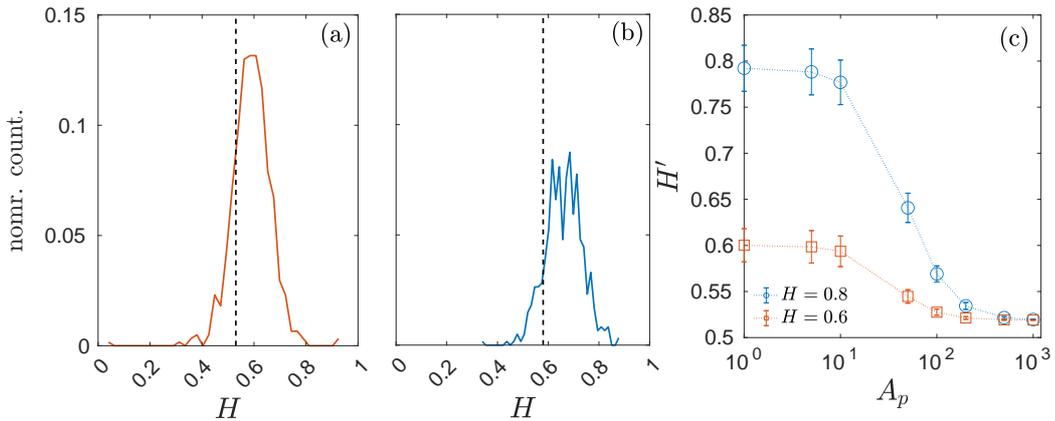}
\caption{Histogram of the estimated Hurst exponents $H$ per day for (a) Frankfurt and (b) Heathrow airports. Dashed black lines indicate $H$ estimated from the entire time series (nights included). (c) Averaged $H$, over 100 independent realisations, as a function of the amplitud of the outliers $A_p$ added to a linear correlated sequence following a power-law distribution with $\alpha=2$ ($N=10^{5}$ and $\tau = 720$).}
\label{fig:H_H_exp}
\end{figure*}

Lastly, we focus on the presence of linear correlations in the landing return intervals per day. We consider Frankfurt and Heathrow airports only, in order to get sequences long enough, and consequently a reliable estimation of $H$. Fig.~\ref{fig:H_H_exp}(a) and (b) show a histogram of the Hurst exponents estimated for the 609 days of Frankfurt and Heathrow airports respectively. We found that the landing return interval dynamics is linearly correlated. By comparing this to the exponent $H$ estimated from the whole data set (nights included), we observe that the presence of outliers underestimate the degree of temporal correlation---see dashed black lines in Fig.~\ref{fig:H_H_exp}(a) and (b). To corroborate this finding, we have simulated linearly correlated data with $H=0.6$ and 0.8 following a power-law distribution with $\alpha=2$, corrupted by additive periodic outliers, with $\tau=720$ and different amplitudes, as described in Sec.~\ref{sec:spikes}. For generating temporal correlated time series with the desired probability distribution, we have followed the iterative algorithm introduced by Schreiber and Schimitz~\cite{schreiber1996improved}. In Fig.~\ref{fig:H_H_exp}(c) is observed that the degree of correlation is underestimated due to the presence of additive periodic outliers. The larger the amplitud $A_p$, the closer is the estimated Hurst exponent to 0.5. Then, we conclude that the intermittency in the landing return interval dynamics due to nights (large and periodic return intervals) leads to a slightly decrease of the linear correlation. From an applied point of view, the dynamics of landing return intervals can be used as a way of characterising the underlying airport dynamics. Specifically, one may hypothesise that the time between successive landings may present correlations when the airport reaches a saturation state, i.e. when the landing of one aircraft is limited by those of the aircraft preceding it. Results here reported indicate that the calculation of a simple $H$ may not be enough, as the presence of outliers and observational noise leads to an underestimation.

%%%%%%%%%%%%%%%%%%%%%%%%%%%%% FIG 3 exp
%%%%%%%%%%%%%%%%%%%%%%%%%%%%%%%%%%%%%%%%%%%%%

\section{Conclusions}
\label{sec:conclusions}

By investigating the performance of MF-DFA applied to finite uncorrelated power-law distributed data in the presence of additive noise and periodic and random-located outliers, we have evidenced that the original bi-fractal properties of the data are considerable corrupted by those artefacts.  In particular, short time series overestimate exponents $h_q$, for negative moments and for $\alpha \in [0.5, 2]$, which leads to a spurious multifractal spectrum rather than a bi-fractral one. This result may be useful for analysing short time series, to avoid erroneous conclusions. When dealing with observational noise, we found that bi-fractality is corrupted for $q<0$. However, for $\alpha=2$ we found a noise-enhanced bi-fractality phenomenon, i.e. the presence of noise offsets the spurious multifractality originated by finite-size effects. This result is valuable for analysing experimental data characterised by $\alpha \sim 2$. The influence of additive periodic outliers corrupts the bi-fractal spectrum around $q=1$, since the outliers affect the moments close to the mean value. This effect does not change the estimation of the multifractal degree, yet it generates an inverted behaviour of the exponent $h_q$ with $q$. Finally, randomly-located outliers deform the entire bi-fractal spectrum, making the theoretical prediction not longer valid for all moments $q$. We believe that the present study gives a more solid characterisation of the bi-fractal nature in empirical time series by MF-DFA approach.

\section*{Acknowledgments}

This project has received funding from the European Research Council (ERC) under the European Union's Horizon 2020 research and innovation programme (grant agreement No 851255).

Financial support has been received from the Agencia Estatal de Investigaci\'on (AEI, MCI, Spain) and Fondo Europeo de Desarrollo Regional (FEDER, UE), under the Maria de Maeztu Program for units of Excellence in R\&D (MDM-2017-0711).

\bibliographystyle{elsarticle-num}
\bibliography{biblio}

\providecommand{\noopsort}[1]{}\providecommand{\singleletter}[1]{#1}%
\begin{thebibliography}{10}
\expandafter\ifx\csname url\endcsname\relax
  \def\url#1{\texttt{#1}}\fi
\expandafter\ifx\csname urlprefix\endcsname\relax\def\urlprefix{URL }\fi
\expandafter\ifx\csname href\endcsname\relax
  \def\href#1#2{#2} \def\path#1{#1}\fi

\bibitem{benzi1984multifractal}
R.~Benzi, G.~Paladin, G.~Parisi, A.~Vulpiani, On the multifractal nature of
  fully developed turbulence and chaotic systems, Journal of Physics A:
  Mathematical and General 17~(18) (1984) 3521.

\bibitem{chhabra1989direct}
A.~B. Chhabra, C.~Meneveau, R.~V. Jensen, K.~Sreenivasan, Direct determination
  of the f ($\alpha$) singularity spectrum and its application to fully
  developed turbulence, Physical Review A 40~(9) (1989) 5284.

\bibitem{meneveau1989measurement}
C.~Meneveau, K.~Sreenivasan, Measurement of ƒ ($\alpha$) from scaling of
  histograms, and applications to dynamical systems and fully developed
  turbulence, Physics Letters A 137~(3) (1989) 103--112.

\bibitem{zhang2019multifractal}
X.~Zhang, H.~Liu, Y.~Zhao, X.~Zhang, Multifractal detrended fluctuation
  analysis on air traffic flow time series: A single airport case, Physica A:
  Statistical Mechanics and its Applications 531 (2019) 121790.

\bibitem{cajueiro2007long}
D.~O. Cajueiro, B.~M. Tabak, Long-range dependence and multifractality in the
  term structure of libor interest rates, Physica A: Statistical Mechanics and
  its Applications 373 (2007) 603--614.

\bibitem{jin2006origins}
H.~Jin, J.~Lu, Origins of the multifractality in shanghai stock market, Nuovo
  Cimento B Serie 121~(9) (2006) 987--994.

\bibitem{zunino2008multifractal}
L.~Zunino, B.~M. Tabak, A.~Figliola, D.~P{\'e}rez, M.~Garavaglia, O.~Rosso, A
  multifractal approach for stock market inefficiency, Physica A: Statistical
  Mechanics and its Applications 387~(26) (2008) 6558--6566.

\bibitem{zhang2019analysis}
X.~Zhang, L.~Yang, Y.~Zhu, Analysis of multifractal characterization of bitcoin
  market based on multifractal detrended fluctuation analysis, Physica A:
  Statistical Mechanics and its Applications 523 (2019) 973--983.

\bibitem{rak2018quantitative}
R.~Rak, D.~Grech, Quantitative approach to multifractality induced by
  correlations and broad distribution of data, Physica A: Statistical Mechanics
  and its Applications 508 (2018) 48--66.

\bibitem{maity2015multifractal}
A.~K. Maity, R.~Pratihar, A.~Mitra, S.~Dey, V.~Agrawal, S.~Sanyal, A.~Banerjee,
  R.~Sengupta, D.~Ghosh, Multifractal detrended fluctuation analysis of alpha
  and theta eeg rhythms with musical stimuli, Chaos, Solitons \& Fractals 81
  (2015) 52--67.

\bibitem{ivanov1999multifractality}
P.~C. Ivanov, L.~A.~N. Amaral, A.~L. Goldberger, S.~Havlin, M.~G. Rosenblum,
  Z.~R. Struzik, H.~E. Stanley, Multifractality in human heartbeat dynamics,
  Nature 399~(6735) (1999) 461--465.

\bibitem{olivares2021high}
F.~Olivares, G.~Funes, D.~G. Perez, High frequency multifractality in return
  intervals from fading induced by turbulence, Fractals 29~(2) (2021)
  2150049--1888.

\bibitem{kantelhardt2003multifractality}
J.~W. Kantelhardt, D.~Rybski, S.~A. Zschiegner, P.~Braun, E.~Koscielny-Bunde,
  V.~Livina, S.~Havlin, A.~Bunde, Multifractality of river runoff and
  precipitation: comparison of fluctuation analysis and wavelet methods,
  Physica A: Statistical Mechanics and its Applications 330~(1-2) (2003)
  240--245.

\bibitem{telesca2005multifractal}
L.~Telesca, V.~Lapenna, M.~Macchiato, Multifractal fluctuations in seismic
  interspike series, Physica A: Statistical Mechanics and its Applications 354
  (2005) 629--640.

\bibitem{telesca2006measuring}
L.~Telesca, V.~Lapenna, Measuring multifractality in seismic sequences,
  Tectonophysics 423~(1-4) (2006) 115--123.

\bibitem{laib2018multifractal}
M.~Laib, J.~Golay, L.~Telesca, M.~Kanevski, Multifractal analysis of the time
  series of daily means of wind speed in complex regions, Chaos, Solitons \&
  Fractals 109 (2018) 118--127.

\bibitem{oswikecimka2011computational}
P.~O{\'s}wiecimka, J.~Kwapie{\'n}, I.~Celi{\'n}ska, S.~Dro{\.z}d{\.z}, R.~Rak,
  Computational approach to multifractal music, arXiv preprint arXiv:1106.2902
  (2011).

\bibitem{ausloos2012generalized}
M.~Ausloos, Generalized hurst exponent and multifractal function of original
  and translated texts mapped into frequency and length time series, Physical
  Review E 86~(3) (2012) 031108.

\bibitem{drozdz2016quantifying}
S.~Dro{\.z}d{\.z}, P.~O{\'s}wiecimka, A.~Kulig, J.~Kwapie{\'n}, K.~Bazarnik,
  I.~Grabska-Gradzi{\'n}ska, J.~Rybicki, M.~Stanuszek, Quantifying origin and
  character of long-range correlations in narrative texts, Information Sciences
  331 (2016) 32--44.

\bibitem{leung2011temporal}
Y.~Leung, E.~Ge, Z.~Yu, Temporal scaling behavior of avian influenza a (h5n1):
  the multifractal detrended fluctuation analysis, Annals of the Association of
  American Geographers 101~(6) (2011) 1221--1240.

\bibitem{kantelhardt2002multifractal}
J.~W. Kantelhardt, S.~A. Zschiegner, E.~Koscielny-Bunde, S.~Havlin, A.~Bunde,
  H.~E. Stanley, Multifractal detrended fluctuation analysis of nonstationary
  time series, Physica A: Statistical Mechanics and its Applications 316~(1-4)
  (2002) 87--114.

\bibitem{grech2013multifractal}
D.~Grech, G.~Pamu{\l}a, On the multifractal effects generated by monofractal
  signals, Physica A: Statistical Mechanics and its Applications 392~(23)
  (2013) 5845--5864.

\bibitem{nakao2000multi}
H.~Nakao, Multi-scaling properties of truncated l{\'e}vy flights, Physics
  Letters A 266~(4-6) (2000) 282--289.

\bibitem{oswiecimka2006wavelet}
P.~O{\'s}wirecimka, J.~Kwapie{\'n}, S.~Dro{\.z}d{\.z}, Wavelet versus detrended
  fluctuation analysis of multifractal structures, Physical Review E 74~(1)
  (2006) 016103.

\bibitem{barabasi1991multifractal}
A.-L. Barab{\'a}si, P.~Sz{\'e}pfalusy, T.~Vicsek, Multifractal spectra of
  multi-affine functions, Physica A: Statistical Mechanics and its Applications
  178~(1) (1991) 17--28.

\bibitem{muzy1994multifractal}
J.-F. Muzy, E.~Bacry, A.~Arneodo, The multifractal formalism revisited with
  wavelets, International Journal of Bifurcation and Chaos 4~(02) (1994)
  245--302.

\bibitem{zhou2008multifractal}
W.-X. Zhou, et~al., Multifractal detrended cross-correlation analysis for two
  nonstationary signals, Physical Review E 77~(6) (2008) 066211.

\bibitem{serrano2009wavelet}
E.~Serrano, A.~Figliola, Wavelet leaders: a new method to estimate the
  multifractal singularity spectra, Physica A: Statistical Mechanics and its
  Applications 388~(14) (2009) 2793--2805.

\bibitem{xiong2017weighted}
H.~Xiong, P.~Shang, Weighted multifractal analysis of financial time series,
  Nonlinear Dynamics 87~(4) (2017) 2251--2266.

\bibitem{jiang2019multifractal}
Z.-Q. Jiang, W.-J. Xie, W.-X. Zhou, D.~Sornette, Multifractal analysis of
  financial markets: a review, Reports on Progress in Physics 82~(12) (2019)
  125901.

\bibitem{kantelhardt2001detecting}
J.~W. Kantelhardt, E.~Koscielny-Bunde, H.~H. Rego, S.~Havlin, A.~Bunde,
  Detecting long-range correlations with detrended fluctuation analysis,
  Physica A: Statistical Mechanics and its Applications 295~(3-4) (2001)
  441--454.

\bibitem{thompson2016multifractal}
J.~R. Thompson, J.~R. Wilson, Multifractal detrended fluctuation analysis:
  Practical applications to financial time series, Mathematics and Computers in
  Simulation 126 (2016) 63--88.

\bibitem{ihlen2012introduction}
E.~A. F.~E. Ihlen, Introduction to multifractal detrended fluctuation analysis
  in matlab, Frontiers in physiology 3 (2012) 141.

\bibitem{drozdz2010quantitative}
S.~Dro{\.z}d{\.z}, J.~Kwapie{\'n}, P.~O{\'s}wiecimka, R.~Rak, Quantitative
  features of multifractal subtleties in time series, EPL (Europhysics Letters)
  88~(6) (2010) 60003.

\bibitem{turiel2006numerical}
A.~Turiel, C.~J. P{\'e}rez-Vicente, J.~Grazzini, Numerical methods for the
  estimation of multifractal singularity spectra on sampled data: A comparative
  study, Journal of Computational Physics 216~(1) (2006) 362--390.

\bibitem{ludescher2011spurious}
J.~Ludescher, M.~I. Bogachev, J.~W. Kantelhardt, A.~Y. Schumann, A.~Bunde, On
  spurious and corrupted multifractality: The effects of additive noise,
  short-term memory and periodic trends, Physica A: Statistical Mechanics and
  its Applications 390~(13) (2011) 2480--2490.

\bibitem{gulich2012effects}
D.~Gulich, L.~Zunino, The effects of observational correlated noises on
  multifractal detrended fluctuation analysis, Physica A: Statistical Mechanics
  and its Applications 391~(16) (2012) 4100--4110.

\bibitem{oswiecimka2020wavelet}
P.~O{\'s}wiecimka, S.~Dro{\.z}d{\.z}, M.~Frasca, R.~Gebarowski, N.~Yoshimura,
  L.~Zunino, L.~Minati, Wavelet-based discrimination of isolated singularities
  masquerading as multifractals in detrended fluctuation analyses, Nonlinear
  Dynamics 100~(2) (2020) 1689--1704.

\bibitem{halsey1986fractal}
T.~C. Halsey, M.~H. Jensen, L.~P. Kadanoff, I.~Procaccia, B.~I. Shraiman,
  Fractal measures and their singularities: The characterization of strange
  sets, Physical review A 33~(2) (1986) 1141.

\bibitem{newman2005power}
M.~E. Newman, Power laws, pareto distributions and zipf's law, Contemporary
  physics 46~(5) (2005) 323--351.

\bibitem{schafer2014bringing}
M.~Sch{\"a}fer, M.~Strohmeier, V.~Lenders, I.~Martinovic, M.~Wilhelm, Bringing
  up opensky: A large-scale ads-b sensor network for research, in: IPSN-14
  Proceedings of the 13th International Symposium on Information Processing in
  Sensor Networks, IEEE, 2014, pp. 83--94.

\bibitem{williams2009gps}
G.~Williams, Gps for the sky: A survey of automatic dependent
  surveillance-broadcast (ads-b) and its implementation in the united states,
  J. Air L. \& Com. 74 (2009) 473.

\bibitem{salcido2017analysis}
R.~Salcido, A.~Kendall, Y.~Zhao, Analysis of automatic dependent
  surveillance-broadcast data, in: 2017 AAAI Fall Symposium Series, 2017.

\bibitem{virkar2012power}
Y.~S. Virkar, Power-law distributions and binned empirical data, Ph.D. thesis,
  University of Colorado at Boulder (2012).

\bibitem{theiler1992testing}
J.~Theiler, S.~Eubank, A.~Longtin, B.~Galdrikian, J.~D. Farmer, Testing for
  nonlinearity in time series: the method of surrogate data, Physica D:
  Nonlinear Phenomena 58~(1-4) (1992) 77--94.

\bibitem{schreiber1996improved}
T.~Schreiber, A.~Schmitz, Improved surrogate data for nonlinearity tests,
  Physical review letters 77~(4) (1996) 635.

\end{thebibliography}

\end{document}